\documentclass[prb,aps,amsmath,amssymb,superscriptaddress,longbibliography,notitlepage,twocolumn]{revtex4-1}
\usepackage{graphicx,multirow,outlines,ulem}
\usepackage{color}
\usepackage{hhline}
\usepackage{physics}
\usepackage{hyperref}

\usepackage[mathscr]{eucal}

\usepackage{tikz}
\usepackage{xcolor}

\newcommand{\bo}{\begin{outline}}
\newcommand{\eo}{\end{outline}}

\newcommand{\qed}{\nobreak \ifvmode \relax \else
      \ifdim\lastskip<1.5em \hskip-\lastskip
      \hskip1.5em plus0em minus0.5em \fi \nobreak
      \vrule height0.75em width0.5em depth0.25em\fi}

\begin{document}

\title{One-dimensional L{\'e}vy Quasicrystal}
\author{Pallabi Chatterjee}
\email{ph22d001@iittp.ac.in}    
\author{Ranjan Modak}
\email{ranjan@iittp.ac.in}    
\affiliation{ Department of Physics, Indian Institute of Technology Tirupati, Tirupati, India~517619} 
\begin{abstract}
Space-fractional quantum mechanics (SFQM) is a generalization of the standard quantum mechanics when the Brownian trajectories
in Feynman path integrals are replaced by L{\'e}vy flights. 
We introduce L{\'e}vy quasicrystal by discretizing the space-fractional Schr$\ddot{\text{o}}$dinger equation using the Gr$\ddot{\text{u}}$nwald-Letnikov derivatives and adding on-site quasiperiodic potential. The discretized version of the usual Schr$\ddot{\text{o}}$dinger equation maps to the 
Aubry-Andr{\'e} Hamiltonian, which supports localization-delocalization transition even in one dimension. 
We find \textcolor{black}{the similarities between 
L{\'e}vy quasicrystal and the Aubry-Andr{\'e} (AA) 
model with power-law hopping, and show 
that the L{\'e}vy quasicrystal supports a delocalization-localization transition as one tunes the quasiperiodic potential strength and shows the coexistence of localized and delocalized states separated by mobility edge. Hence, a possible realization of SFQM in optical experiments should be a new experimental platform to test the predictions of AA models in the presence of power-law hopping.}
\end{abstract}
\maketitle

\section{Introduction}
In usual quantum mechanics, the typical energy-momentum relation for a particle of mass $m$ is given by $E=P^2/2m$.
In general, one can have a situation where the energy-momentum relation is given by $E\propto P^{\alpha}$~\cite{PLADARTORA2021127643}, where $\alpha$ can be a fraction. This kind of situation can be described by the Space Fractional Quantum Mechanics(SFQM), which was introduced by Laskin~\cite{1LASKIN2000298,2LaskinPhysRevE.62.3135,3LaskinPhysRevE.66.056108,levy2013PhysRevE.88.012120,hasan2018tunneling,zhang.2015}. SFQM is a natural generalization of the standard
quantum mechanics that arises when the Brownian trajectories
in Feynman path integrals are replaced by L{\'e}vy flights. The
classical L{\'e}vy flight is a stochastic process that, in one
dimension is described by a jump length probability density
function (PDF) of the form, $\Pi_\alpha(x)\propto1/|x|^{\alpha+1}$ for $|x|\to \infty$, with $\alpha\in(0,2]$ \cite{levy2013PhysRevE.88.012120}, where
$\alpha$ is known as L{\'e}vy index.

There have been numerous applications of classical L{\'e}vy flights. Especially, in the context of 
successfully predicting the anomalous scaling of dynamical correlations of conserved quantities in one-dimensional (1D) Hamiltonian systems, the L{\'e}vy scaling for the spreading of
local energy perturbation has been predicted, as well as diverging thermal conductivity (via GreenKubo formula)~\cite{mendl2015current,henk.2012,dhar2013exact,kundu2019fractional}.
Also, in order to understand the motion of the particle in a rotating flow ~\cite{solomon.1993,vikash.2016} or even the traveling behavior of animals~\cite{brockmann2006scaling,benhamou2007many,levy_appl_1,levy_appl_2}, the complex dynamics of real-life financial markets~\cite{yarahmadi20222d},
L{\'e}vy description has been extremely useful. It has been shown recently in Ref.~\cite{levy2013PhysRevE.88.012120}, one can discretize Space fractional Schr$\ddot{\text{o}}$dinger equation, introduce 
a system which is referred to as L{\'e}vy Crystal. Given the extraordinary advancements of ultra-cold experiments in the last two decades, L{\'e}vy Crystal is a potential candidate for an experimentally accessible realization of SFQM in a condensed-matter environment~\cite{kaufman2016quantum, rmp_localization,schreiber2015observation}, however, it has not 
been explicitly demonstrated how that can be achieved in Ref.~\cite{levy2013PhysRevE.88.012120}. \textcolor{black}{
Moreover, fractional quantum mechanics plays a very crucial role in optical systems
~\cite{op1,op2,op3,op4,huang2019localization,min.2015}.  In this context, there have been recent theoretical works on domain walls in fractional media~\cite{domainwall}, fractional diffraction in the context of parity-time symmetric potentials ~\cite{fracpt1,fracpt2,fracpt3}, and on optical solitons~\cite{soliton44,soliton43,soliton42,soliton40,soliton39,soliton38,soliton37,soliton36,soliton35,soliton34,soliton33,soliton32,soliton31,soliton30,soliton29,soliton28,soliton27,soliton26,soliton25,soliton24,soliton22,soliton21,soliton20}, that bought lots of attention. The experimental realization of an optical system representing the fractional Schr$\ddot{\text{o}}$dinger equation has been reported very recently in Ref.~\cite{liu2023experimental}. 
Also, it is important to point out that in the context of Sisyphus cooling, fractional equations arise from the dynamics of cold atoms, an effect that is well-studied \cite{afek2021anomalous,zoller.96}.}
\begin{figure}
    \centering
    \includegraphics[width=0.49\textwidth,height=0.33\textwidth]{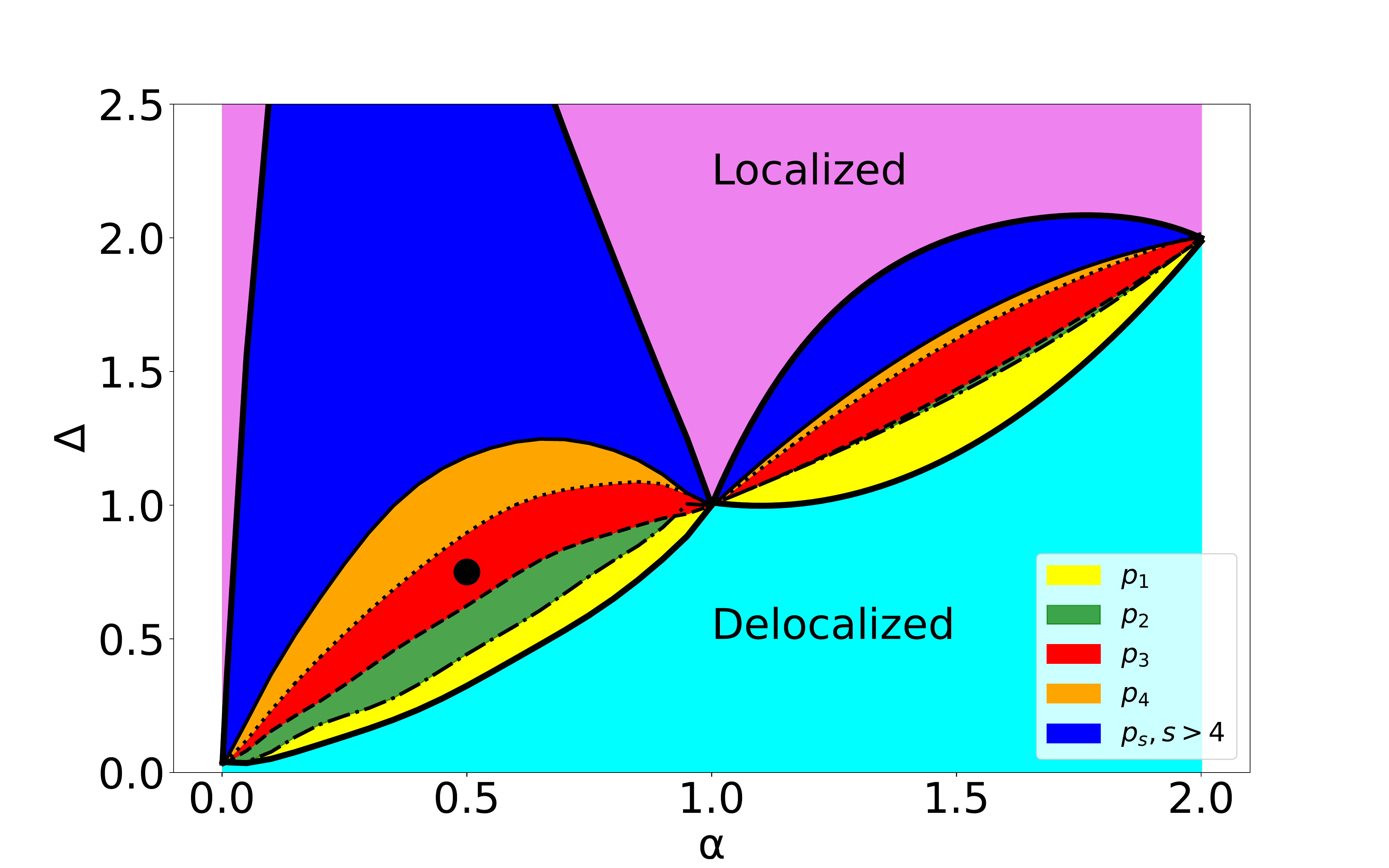}
    \caption{Here we show the phase diagram of the L{\'e}vy
    quasicrystal. Black circle represents $\alpha=0.5$ and $\Delta=0.75$. In the main text, we show the results corresponding to this point.}
    \label{fig0}
\end{figure}

On the other hand, the disorder is ubique in nature, and in the condensed matter system, it plays a very important role. In the one-dimensional(1D) system, the disorder has an extreme consequence, even a tiny bit of disorder is sufficient 
to localize all the single-particle states. This phenomenon is famously known as Anderson localization ~\cite{tvr.1979,tvr.1985,anderson.1958}. In recent days, there 
has been a plethora of work to understand the interplay of 
interaction and disorder, and that cause a delocalization-localization transition in the many-body Fock space, which is referred to as many-body localization (MBL) transition~\cite{rmp_localization,moore.2012,vosk.2013,basko2006metal,smukerjee.2018,smukerjee.2022}. In the experimental point of view, realizing true disorder in ultra-cold experiments is non-trivial, hence one of the obvious candidates is to replace it with the quasiperiodic potential. In contrast to the true disorder, quasiperiodic potential can cause a delocalization-localization transition in 1D even in the absence of interaction, and this Hamiltonian is known as Aubry-Andr{\'e} (AA) model~\cite{aubry.1980}. 
\textcolor{black}{
While the nature of the localized phase observed due to quasiperiodic potentials and true disorders are the same (in both cases, eigenstates are exponentially localized), there are differences
in critical properties associated with the localization-delocalization transition. 
In both cases, the localization length $\xi$ diverges at the transition following as $\xi\sim \delta ^{-\nu}$, where $\delta$ is the distance to the critical point in the parameter space and $\nu$ is the localization length exponent. In true disorder-driven localization,  there is a rigorous bound on the localization length exponent $\nu$, i.e.,  it must satisfy $\nu \ge 2/d$ criteria to ensure the stability of the transition~\cite{ccfs}. However, in quasiperiodic models, such criteria do not apply. For most of the one-dimensional quasi-periodic models, $\nu$ is close to $1$~\cite{IPRdeng2017many}.}
In recent days, there have been numerous studies, both theoretically and experimentally, involving a variant of the AA model that shows the coexistence of both localized and delocalized states separated by a mobility edge ~\cite{qp1,qp2,qp4,qp3,qp5,qp6,qp7,archak.2017,archak.2018,modak.2018,qp70,modak.dqpt,add_qp1,add_qp2,add_qp3}.

In this work, one of the main aims is to address the question of what happens to the fate of Anderson localization in the context of SFQM in the presence of quasiperiodic potential, we call it L{\'e}vy quasicrystal. 
\textcolor{black}{Interestingly, a similar question has been asked in a recent study~\cite{huang2019localization}, but a detailed theoretical understanding of the phase diagram of the different phases was lacking there. Also, the L{\'e}vy-flight models have been used extensively in understanding real-life financial markets, the traveling behavior of humans,  and even biological systems. If one wants to model such a system and take into account correlated random events, our L{\'e}vy quasicrystal model can be an extremely suitable candidate for such cases.}
We demonstrate the phase diagram as a function of  $\alpha$ and the strength of the quasiperiodic potential $\Delta$, which is shown in Fig.~\ref{fig0}. While we find that like the Aubry-Andr{\'e} model (which is the discretized version of the usual Schr$\ddot{\text{o}}$dinger equation), there exists a completely delocalized (DL) and Anderson localized (AL) phase, but on top of that there is a parameter region where these two types of states co-exist, and they are separated by the mobility edge (ME). Those ME phases can also be characterized as different $p_s$ phases (where $s>0$ can be any positive integer) based on the fraction of delocalized states in the spectrum.  
\textcolor{black}{We also compare our model with a potential experimentally realizable AA model with power-law hopping.}

The paper is organized as follows. In Sec. II, we introduce
the model of L\'evy quasicrystal. Next, we discuss the analytical prediction of the mobility edge in Sec. III. In Sec. IV, we show our numerical results. Sec. V shows the comparison between L\'evy quasicrystal and the effective power-law hopping model and finally, we summarize our results in Sec. VI. 

\section{Model}
The one-dimentional space-fractional Schr$\ddot{\text{o}}$dinger equation is given by,
\begin{eqnarray}
  H|\psi\rangle=D_\alpha P^\alpha|\psi\rangle + V|\psi\rangle=E|\psi\rangle ,
\end{eqnarray}
where $D_\alpha \in \mathbb{R}$ is a constant (note that $D_2$ is equivalent of inverse of mass term in usual Schr$\ddot{\text{o}}$dinger equation),
$P^\alpha$ is the $\alpha$-th power of the momentum operator,
$V$ is the potential energy operator and $|\psi\rangle$ is the eigenstate with eigenvalue $E$
\cite{1LASKIN2000298,2LaskinPhysRevE.62.3135,levy2013PhysRevE.88.012120}. 

The position space representation of $\alpha$-th power of the  momentum operator is  $\langle x|P^\alpha|\psi\rangle=-\hbar^\alpha \mathcal{D}^{\alpha}_{|x|} \psi(x)$ ~\cite{3LaskinPhysRevE.66.056108,levy2013PhysRevE.88.012120}, 
where $\mathcal{D}_{|x|}^{\alpha}$ is the Riesz Fractional Derivative of order $\alpha$. While usually to get the finite first moment of the L\'evy process, $\alpha$ is taken within the limit $\alpha \in (1,2]$ ~\cite{3LaskinPhysRevE.66.056108}, 
\textcolor{black}{but
given there exist, biological models, where the L\'evy parameter is taken to be $0<\alpha<1$~\cite{benhamou2007many}, also in a very recent optical experiment, the parameter range $\alpha<1$ of the fractional Schr$\ddot{\text{o}}$dinger equation has been realized~\cite{liu2023experimental} in the temporal domain, that motivates us to even explore the extended parameter regime, and} 
we use the same position space representation of $\alpha$-th power of the  momentum operator for $\alpha \in (0,2]$ to get the discretized version. Riesz Fractional Derivative of order $\alpha$ (note that we consider $\alpha=1$ point separately in the appendix), can be written as~\cite{lavyGorenflo98randomwalk,levy2013PhysRevE.88.012120}(\textcolor{black}{note that the expression of the Riesz Fractional Derivative is taken from Ref.~\cite{lavyGorenflo98randomwalk}}),
\begin{eqnarray}
\label{eqn:3}
 \mathcal{D}^{\alpha}_{|x|}=-\frac{1}{2\cos(\frac{\alpha\pi}{2})}(I_{+}^{-\alpha} + I_{-}^{-\alpha}),
\end{eqnarray}
where $I_{\pm}^{-\alpha}$ are given by  (approximating Gr$\ddot{\text{u}}$nwald-Letnikov operators)~\cite{lavyGorenflo98randomwalk,levy2013PhysRevE.88.012120}, 
\begin{eqnarray}
\label{eqn:4}
I_{\pm}^{-\alpha}\psi(x)=\lim_{a\to0}\frac{1}{a^\alpha}\sum_{n=0}^{\infty}(-1)^n\binom{\alpha}{n}\psi[x\mp an],
\end{eqnarray}
for $ 0<\alpha\le1$.
\begin{eqnarray}
\label{eqn:5}
I_{\pm}^{-\alpha}\psi(x)=\lim_{a\to0}\frac{1}{a^\alpha}\sum_{n=0}^{\infty}(-1)^n\binom{\alpha}{n}\psi[x\mp (n-1)a],
\end{eqnarray}
for $1<\alpha\le2$. Where $a$ stands for a small positive step length. Here, $\binom{\alpha}{n}=\frac{\Gamma(\alpha+1)}{\Gamma(n+1)\Gamma(\alpha-n+1)}$,
where $\Gamma(.)$ is the usual Gamma function.
It is straightforward to check that for $\alpha=2$ Riesz Fractional Derivative becomes the standard second-order derivative.\\

In this work, we consider the potential to be space-dependent (quasiperiodic potential), hence the model can be represented by the equation,
\begin{eqnarray}
\label{new eqn}
\langle x_l| H|\psi\rangle = \langle x_l|D_{\alpha}P^{\alpha}|\psi \rangle + \Delta \cos(2\pi\beta l + \phi)\psi(x_l).
\end{eqnarray}
$x_l$ is equally spaced grid points i.e $x_l=la$, $a$=lattice constant, $a>0 $ and $l\in\mathbb{Z}$.$\beta$ is an irrational number.  We choose $\beta=\frac{\sqrt{5}-1}{2}$, and $\phi$ is a random number chosen between $[0,2\pi]$. 
\textcolor{black}{Note that in the previous study~\cite{levy2013PhysRevE.88.012120} the potential energy term was considered to be periodic in lattice spacing $a$, the construction of the space-fractional Schr$\ddot{\text{o}}$dinger equation via Riesz Fractional Derivative is not only limited to
periodic potential.}
\\

Replacing the exact momentum 
operator  by its discretized version\cite{datta_1995,levy2013PhysRevE.88.012120} and using 
Eqn.~\eqref{eqn:5}, one gets 
for $1<\alpha\le2$,
\begin{multline}
\langle x_l|D_{\alpha}P^{\alpha}|\psi \rangle:=\frac{t_0}{2}\sum_{n=0}^{\infty}(-1)^n\binom{\alpha}{n}\\
(\psi[x_l+(n-1)a]+\psi[x_l-(n-1)a]),
\end{multline}
where $t_0=\frac{D_\alpha \hbar^\alpha}{a^\alpha \cos(\frac{\alpha\pi}{2})}$.

Then, $\langle x_l|D_{\alpha}P^{\alpha}|\psi \rangle$ becomes,\\
\begin{multline}
\langle x_l|D_{\alpha}P^{\alpha}|\psi \rangle:=\\
\frac{t_0}{2}\sum_{n\ne{0}}^{\infty}(-1)^{(n+1)}\binom{\alpha}{n+1}\\
\psi(x_l+ na)+\frac{t_0}{2}[\psi(x_l-a)+\psi(x_l+a)]-\alpha t_0\psi(x_l).
\end{multline}

Next, we drop the constant diagonal term $\alpha t_0 $ (as it will create only a shift to the energy level). \\
Eqn:\eqref{new eqn} becomes,
\begin{eqnarray}
\langle x_l |H|\psi\rangle=\sum_{n\ne{0}}^{\infty}t(n)\psi(x_l+na)+ \Delta \cos(2\pi\beta l + \phi)\psi(x_l),
\end{eqnarray}
where $t(n)$ is the hopping parameter for $1<\alpha\le{2}$,
\begin{eqnarray}
\label{eqn:9}
t(n)=\frac{t_0}{2}[(-1)^{|n|+1}\binom{\alpha}{|n|+1}+\delta_{|n|,1}],
\end{eqnarray}
where $\delta_{n,m}$ is the Kronecker delta.\\

Now, one can repeat similar calculations for $0<\alpha<{1}$, and gets the following equation,
\begin{multline}
\langle x_l|D_{\alpha}P^{\alpha}|\psi \rangle:=\frac{t_0}{2}\sum_{n=0}^{\infty}(-1)^n\binom{\alpha}{n}\\
(\psi[x_l+na]+\psi[x_l-na]).
\end{multline}
Then, $ \langle x_l|D_{\alpha}P^{\alpha}|\psi \rangle$ becomes\footnote{Eqn. 2-11 have already been derived in Ref \cite{levy2013PhysRevE.88.012120} and Ref:\cite{lavyGorenflo98randomwalk}. We have incorporated them in our paper for the self-consistency 
and to improve readability.},\\
\begin{multline}
\langle x_l|D_{\alpha}P^{\alpha}|\psi \rangle=\\
\frac{t_0}{2}\sum_{n\ne{0}}^{\infty}(-1)^{n}\binom{\alpha}{n}\psi(x_l+ na)+t_0\psi(x_l).
\end{multline}

Once again,  subtracting the constant diagonal terms, one gets,
\begin{multline}
    \langle x_l|H|\psi \rangle=\\
    \sum_{n\ne{0}}^{\infty}t(n)\psi(x_l+na) +  \Delta \cos(2\pi\beta l + \phi)\psi(x_l),
\end{multline}

where $t(n)$, the hopping parameter for $0<\alpha<{1}$ is given by,
\begin{eqnarray}
\label{eqn:13}
t(n)=\frac{t_0}{2}[(-1)^{|n|}\binom{\alpha}{|n|}].
\end{eqnarray}

Hence in general our model can be written as,
\begin{eqnarray}
{H}&=&\sum_{j,n\neq0}(t(n){c}^{\dag}_j{c}_{j+n}+\text{H.c.})+\Delta\sum_{j} \cos(2\pi\beta j+\phi){n}_j, \nonumber \\
\label{nonint_model}
\end{eqnarray}
where, ${c}^{\dag}_j$  (${c}_j$) is the fermionic creation (annihilation) operator at site $j$, ${n}_j ={c}^{\dag}_j{c}_{j}$ is the number operator,\\
\begin{eqnarray}
\label{eqn:13}
t(n)&=&-\frac{1}{2}[(-1)^{|n|}\binom{\alpha}{|n|}] ~~~~~~~~~~~~~~~~~~~0<\alpha<{1},\nonumber\\
&=&\frac{i}{2}~~ \delta_{|n|,1}~~~~~~~~~~~~~~~~~~~~~~~~~~~~~~~~~~~\alpha=1,\nonumber\\
&=&\frac{1}{2}[(-1)^{|n|+1}\binom{\alpha}{|n|+1}+\delta_{|n|,1}] ~~~1<\alpha\le{2}. \nonumber\\
\end{eqnarray}
We do average over $\phi$ to obtain better statistics and for all of our calculations, we have used periodic boundary conditions.  \\
For $\alpha=2$, the Hamiltonian ${H}$ is the same as the  Aubry-Andr{\'e} (AA) Hamiltonian, which supports a  delocalization-localization transition as one tunes $\Delta$. In the thermodynamic limit, $\Delta=2$ corresponds to the transition point \cite{aubry.1980}  between localized and delocalized phases and for $\Delta<2$ ($\Delta>2$), all the eigenstates of the model are delocalized (localized). 

\section{Analytical prediction}
Unlike the AA model, the Hamiltonian we are interested in here
Eqn.~\eqref{nonint_model}, which has higher order hopping terms. It has been shown in Ref.~\cite{SDsharma2}, that in the presence of incommensurate potential with long-range hopping, there exists a mobility edge. One example of such a model is the exponential hopping model (where the hopping amplitudes fall off exponentially with increasing the range $n$ of hopping as $te^{-p|n|}$). This exponential model is self-dual \cite{SDsharma2}. One can obtain the mobility edge line, using the self-duality condition of the exponential hopping Hamiltonian, and it is given by the following expression \textcolor{black}{which is taken from Ref.~\cite{SDsharma2}},
\begin{eqnarray}
\label{duality condition1}
\cosh(p)=\frac{E+t}{\Delta}.
\end{eqnarray}
Even though the above expression of ME line is for the self-dual (exponential hopping) model, it has been argued in Ref.~\cite{SDsharma2} that even the non-dual models (e.g. $t_1-t_2$ model, Gaussian hopping model, inverse power law hopping model) can be approximated as an exponential model and the analytical mobility edge line~\eqref{duality condition1} gives the qualitatively correct interpretation, if the falling off of the long-range hopping terms is fast enough. However, in general, the energy-dependent mobility edges are not linear in $\Delta$ in the case of a non-dual model.

In our case, for the Hamiltonian \eqref{nonint_model},  the long-range hopping terms fall off quickly (see Eqn:\eqref{eqn:13}).
Given the decay is fast enough, that inspires us to approximate the ME line using the expression \eqref{duality condition1}, and we choose $p=\ln{\frac{t(1)}{t(2)}}$ and $t=t(1)e^p$~\cite{SDsharma2},  
where $t(1)$ and $t(2)$ are the nearest neighbor and next nearest neighbor hopping terms, respectively.
Next, we will compare our analytic prediction
with the numerical results for the finite-size systems.

\begin{figure}
    \centering
    \includegraphics[width=0.53\textwidth, height=0.3\textwidth]{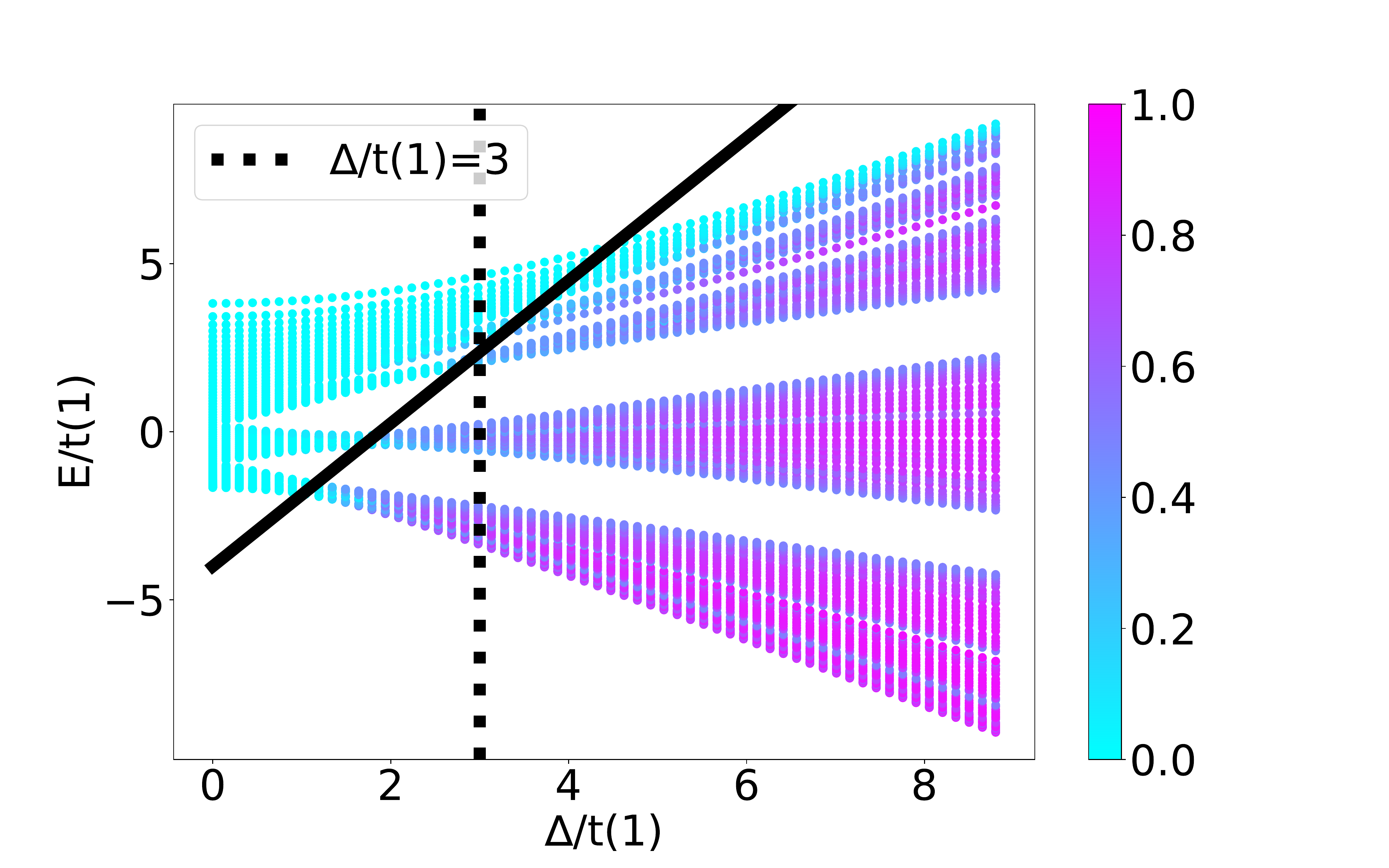}\\
    \includegraphics[width=0.53\textwidth, height=0.3\textwidth]{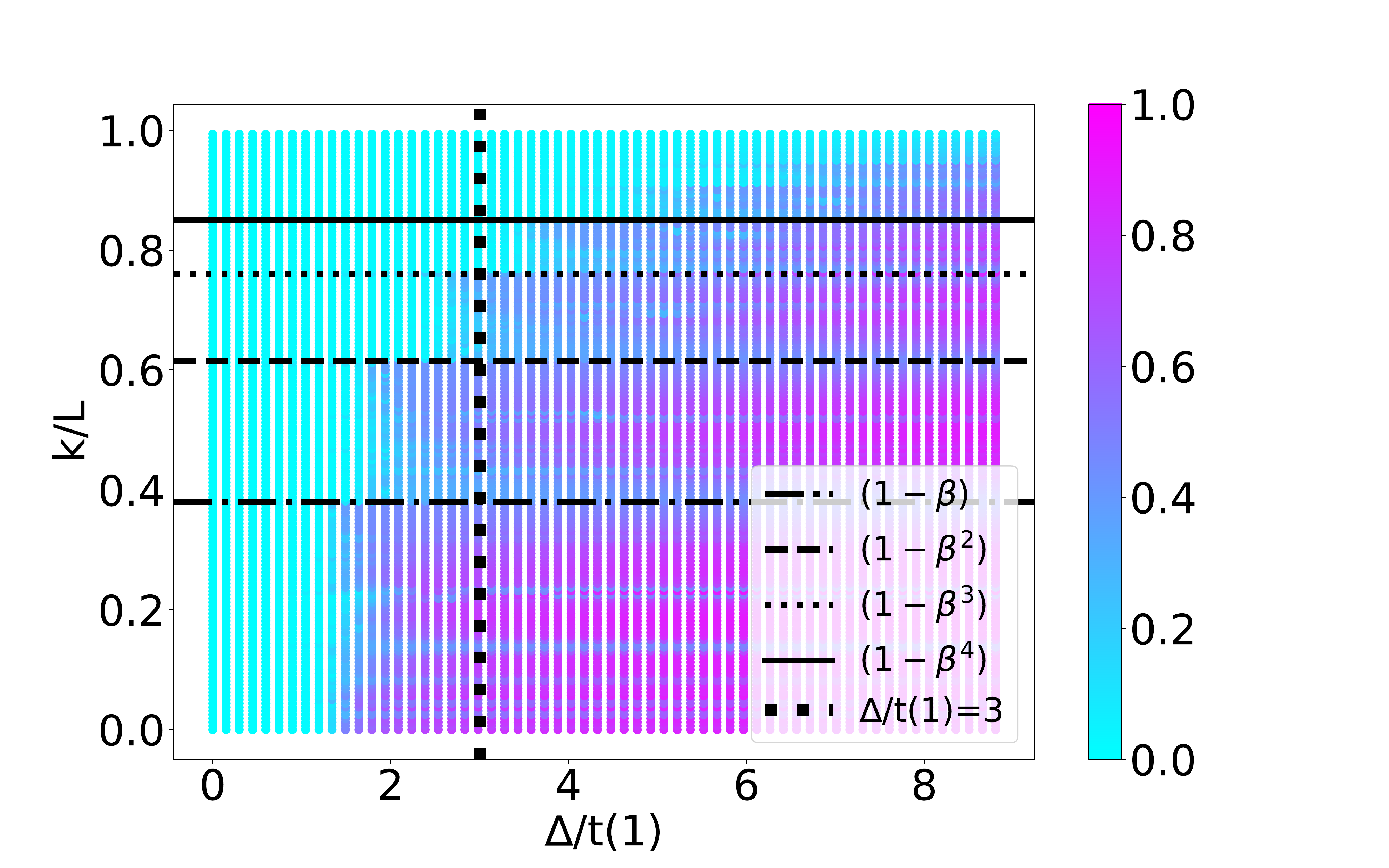}\\
    \includegraphics[width=0.48\textwidth, height=0.3\textwidth]{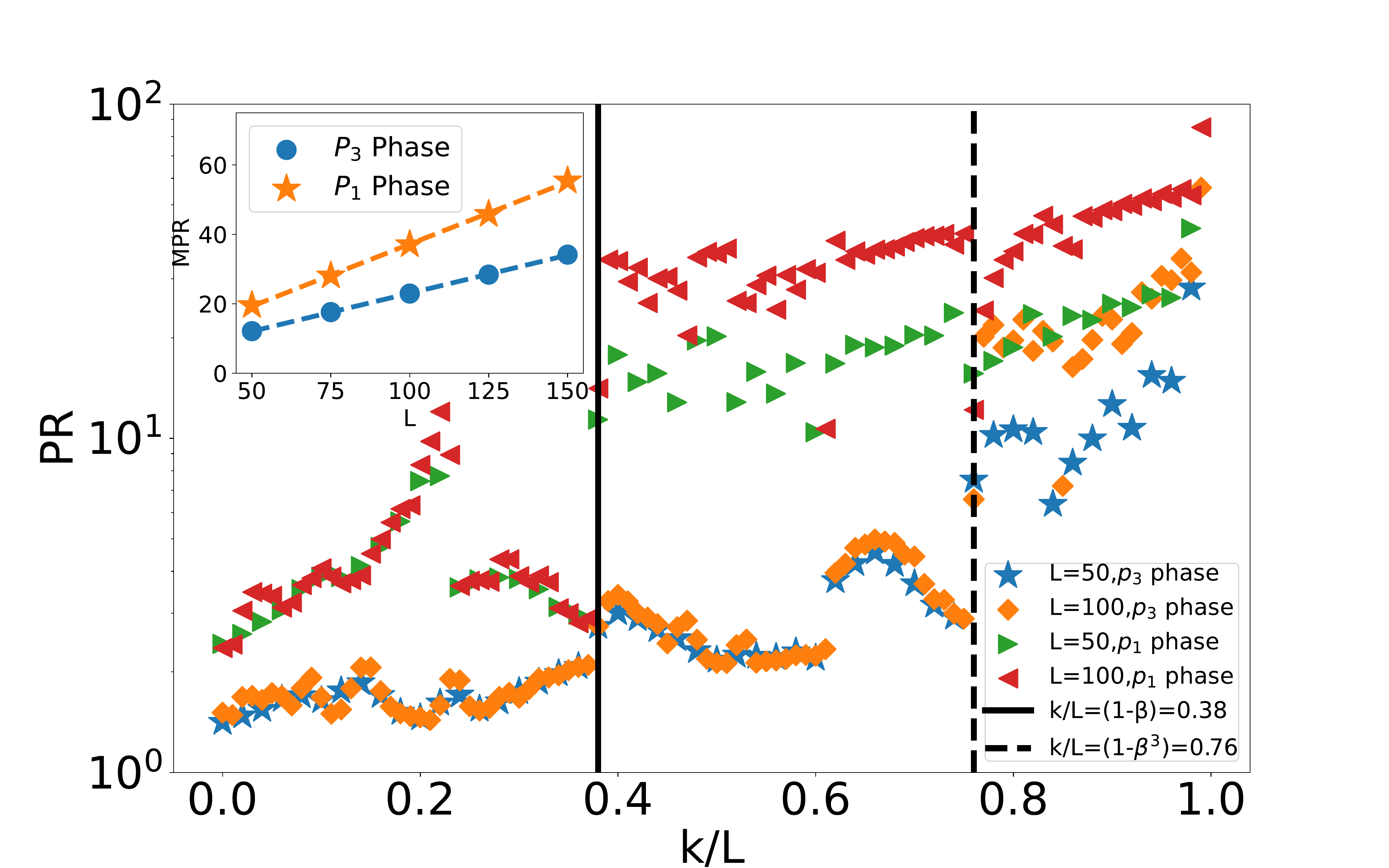}
    \caption{(Upper panel) Contour plot of Energy dependent IPR vs $\Delta/t(1)$ for the Hamiltonian \eqref{nonint_model}.
    The solid line shows the analytical separation between delocalized and localized states. (Middle panel) Shows the 
    variation of IPR with respect to energy level index $k/L$. Results are for $\alpha=0.5$
    and $L=160$.
    \textcolor{black}{(Lower panel) Shows the variation of PR with respect to energy level index $k/L$ of $p_1$ and $p_3$ phases for different system sizes($L=50$ and $L=100$). Inset shows for both $p_1$ and $p_3$ phases, the mean PR value for the delocalized states scales linearly with system size $L$.
    }} 
    \label{fig2}
\end{figure}

\begin{figure}
    \centering
    \includegraphics[width=0.48\textwidth, height=0.3\textwidth]{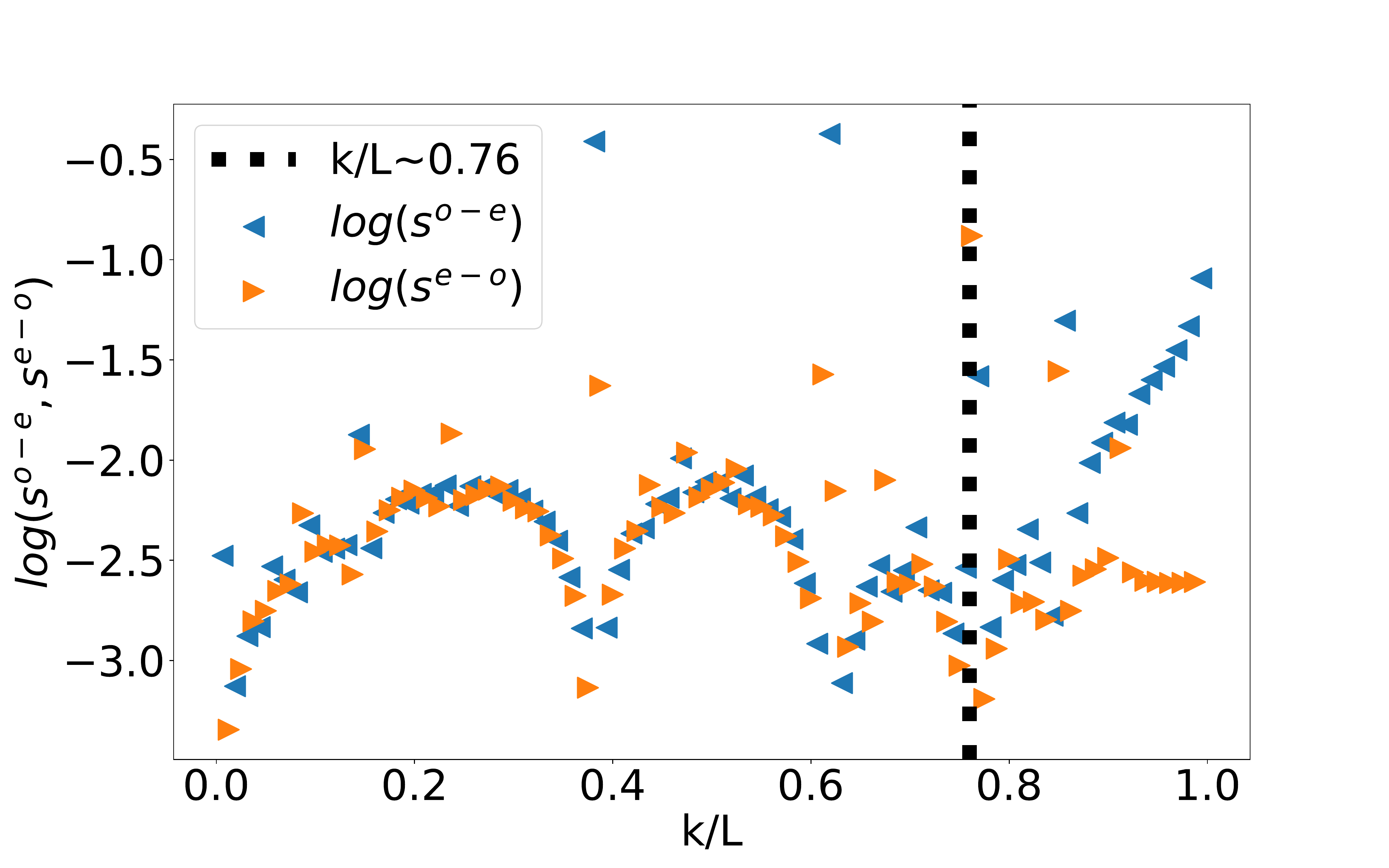}
    \caption{
    Shows the variation of level spacing $s^{e-o}$ and  $s^{o-e}$ vs $k/L$. Results are for $\alpha=0.5$
    and $L=160$.}
    \label{fig2f}
\end{figure}

\section{Numerical Results}
Next,  we report our numerical findings. One of the most popular diagnostic tools to numerically characterize localized (delocalized) eigenstate is to investigate the Inverse Participation Ratio (IPR). IPR for a normalized  eigenstate ($k$-th) is given by~\cite{IPRdeng2017many} $\text{IPR}(k)=\sum_{i=1} ^{L}|{{\psi_i}(k)|}^4$. For a localized state, IPR approaches its maximum possible value i.e. $1$, and for a delocalized state, the IPR value is of the order of $1/L$, which tends to zero in the thermodynamic limit. 

Hence, the behavior of the IPR across the mobility edge line is expected to differ, given ME line separates localized and delocalized states in the spectrum. Figure.~\ref{fig2} (upper panel) shows the variation of energy-dependent IPR as a function of $\Delta/t(1)$ for $\alpha=0.5$. The solid line is the analytic 
prediction of the ME line obtained from Eqn.~\eqref{duality condition1}. Note that even though our analytical prediction 
is for the self-dual model (exponential hopping Hamiltonian), with an appropriate choice of $p$, one can see that it gives  reasonably well qualitative agreement  with our finite-sized 
numerical results for the Hamiltonian $H$. However, we can see, as expected here the mobility edge separation is not exactly linear in $\Delta$. Figure.~\ref{fig2}(middle panel) shows the variation of the IPR with respect to the energy index $k/L$.
This is a signature of the existence of the ME phase consisting of phases where $\beta^s$ fraction of states are DL (rest are AL), where $s>0$ can be any positive integer. We refer to them as $p_s$ ME phases. 
\textcolor{black}{Next, we also investigate the finite size scaling of participation ratio (PR) for eigenstates. Note that PR is essentially the inverse of IPR, which gives an intuitive measure of localization length. If PR does not scale with system size $L$, that is a signature of exponentially localized states, where the localization length is independent of system size. On the other hand, PR for delocalized states is expected to increase with $L$ linearly. Figure.~\ref{fig2} (lower panel) confirms that for both for $p_1$ and $p_3$ phases indeed $\beta$
and $\beta^3$ fraction of states are delocalized. 
Moreover, the inset shows that the mean PR for the delocalized states scales linearly with $L$, 
 which excludes the possibility of having multifractal states.  (Note, the 
scaling of the participation ratio is sub-linear in $L$ for the multifractal phase~\cite{even-odd-levelspacingPhysRevLett.123.025301}.) }

We also strengthen our finding from the even-odd and odd-even level spacing statistics \cite{even-odd-levelspacingPhysRevLett.123.025301}.
Even-Odd and odd-Even spacing are given by,
$s_k^{e-o}=E_{2k}-E_{2k-1}$, $s_k^{o-e}=E_{2k+1}-E_{2k}$,  
where ${E_k}$s are the energy eigenvalues of the single-particle spectrum arranged in ascending order.
For the delocalized single-particle states, there exist almost doubly degenerate spectra \cite{aubry.1980,even-odd-levelspacingPhysRevLett.123.025301}, so there is a gap between $s_k^{e-o}$ and $s_k^{o-e}$. On the other hand, for the localized single-particle states, that degeneracy is lifted, hence the gap between $s_k^{e-o}$ and $s_k^{o-e}$ vanishes. Figure.~\ref{fig2f} clearly 
demonstrates that. We choose $\Delta/t(1)=3$ [which denoted by dashed line in Fig.~\ref{fig2} (upper panel)]. It is clear from
Fig.~\ref{fig2} (upper panel), for this choice of parameter, the single-particle spectrum has a ME (which corresponds to $k/L=0.76\simeq 1-\beta^3$; $p_3$ ME phase). Figure.~\ref{fig2f} shows clearly that $s^{e-0}$ and $s^{o-e}$ fall on top of each other
for $k/L<0.76$ and for $k/L>0.76$ there is a gap, which indicates the existence of the localized and delocalized phase respectively. We identify this point with a black circle in Fig.~\ref{fig0}. 
Also, it is important to point out that in 
the multifractal phase, $s^{e-0}$ and $s^{o-e}$ both are strongly scattered, we don't see any evidence of that in our results. Results for $\alpha=1.5$ are shown in Fig:~\ref {fig5b}. We repeat our analysis for different values of $\alpha$ and obtain the complete phase diagram which is shown in Fig.~\ref{fig0} (see Appendix:\ref{appendixIII} for details).

\begin{figure}
\centering
    \includegraphics[width=0.48\textwidth, height=0.3\textwidth]{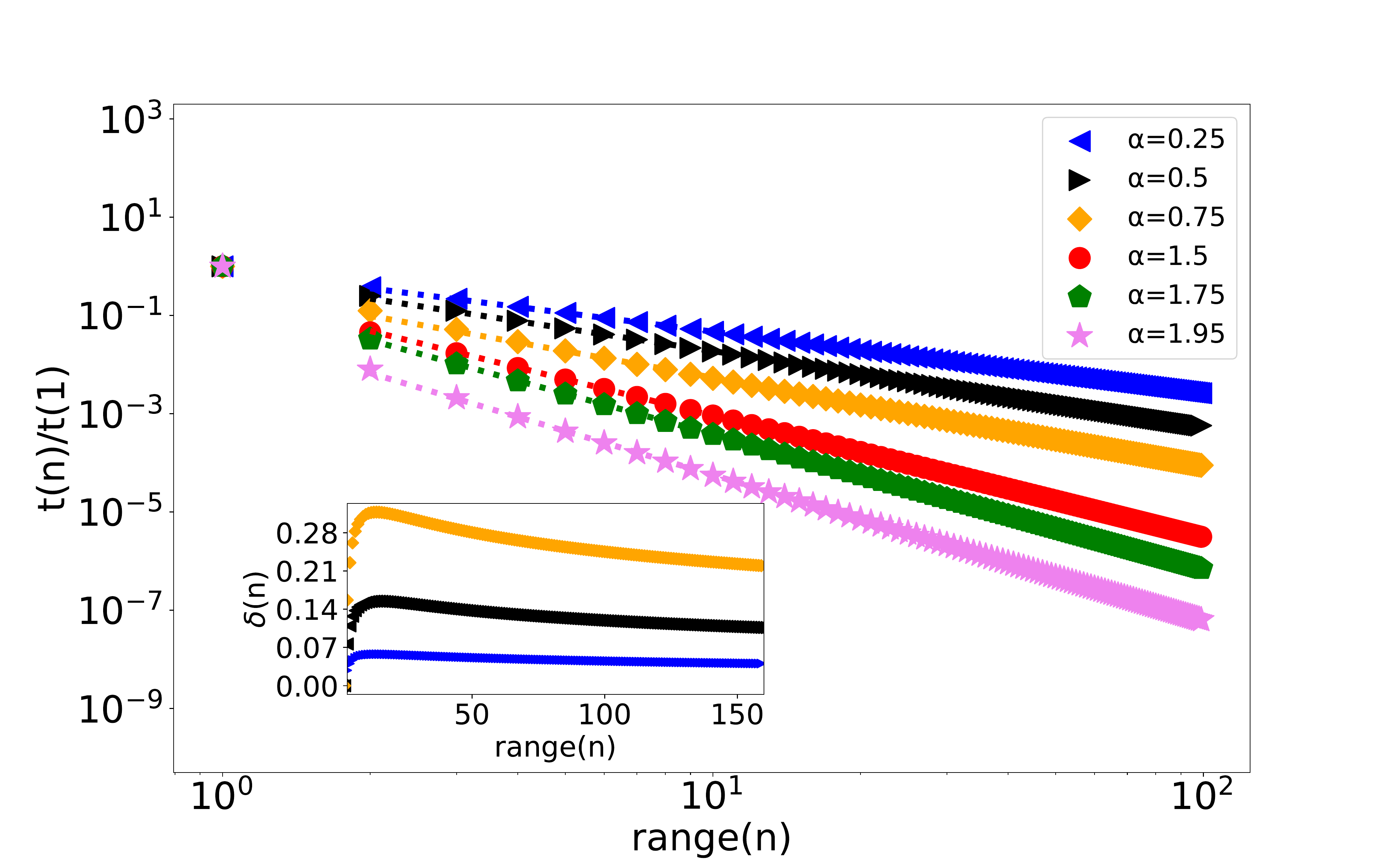}
     \includegraphics[width=0.48\textwidth, height=0.3\textwidth]{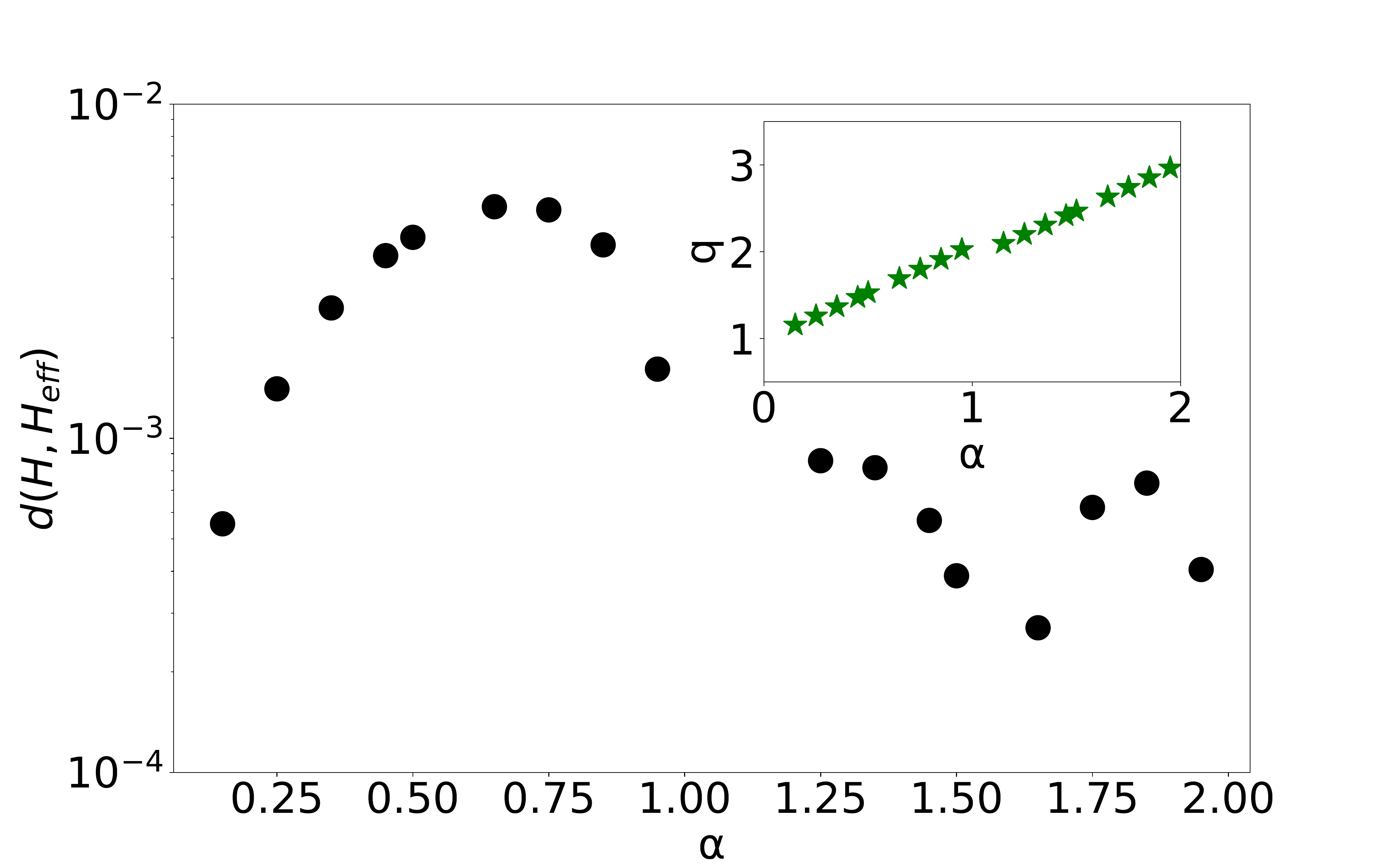}
    \caption{Upper panel: Variation of $t(n)/t(1)$ vs range $n$ for $L=100$
    for different values of $\alpha$. Dashed lines show the power-law fit for $n\ge2$. Inset shows the variation of the difference between the hopping terms coming from the Hamiltonian \eqref{nonint_model} and $H_ {eff}$ normalized by the hopping terms coming from the Hamiltonian \eqref{nonint_model}(denoted by $\delta$) with range $n$. \\
     Lower panel: The normalized trace distance between 
    $H$ and $H_{\text{eff}}$ as a function of $\alpha$. The inset shows the variation $q$ with $\alpha$.}
    \label{fig3}
\end{figure}

 \begin{figure}
    \centering
    \includegraphics[width=0.53\textwidth, height=0.3\textwidth]{fig2a.pdf}
    \includegraphics[width=0.53\textwidth, height=0.3\textwidth]{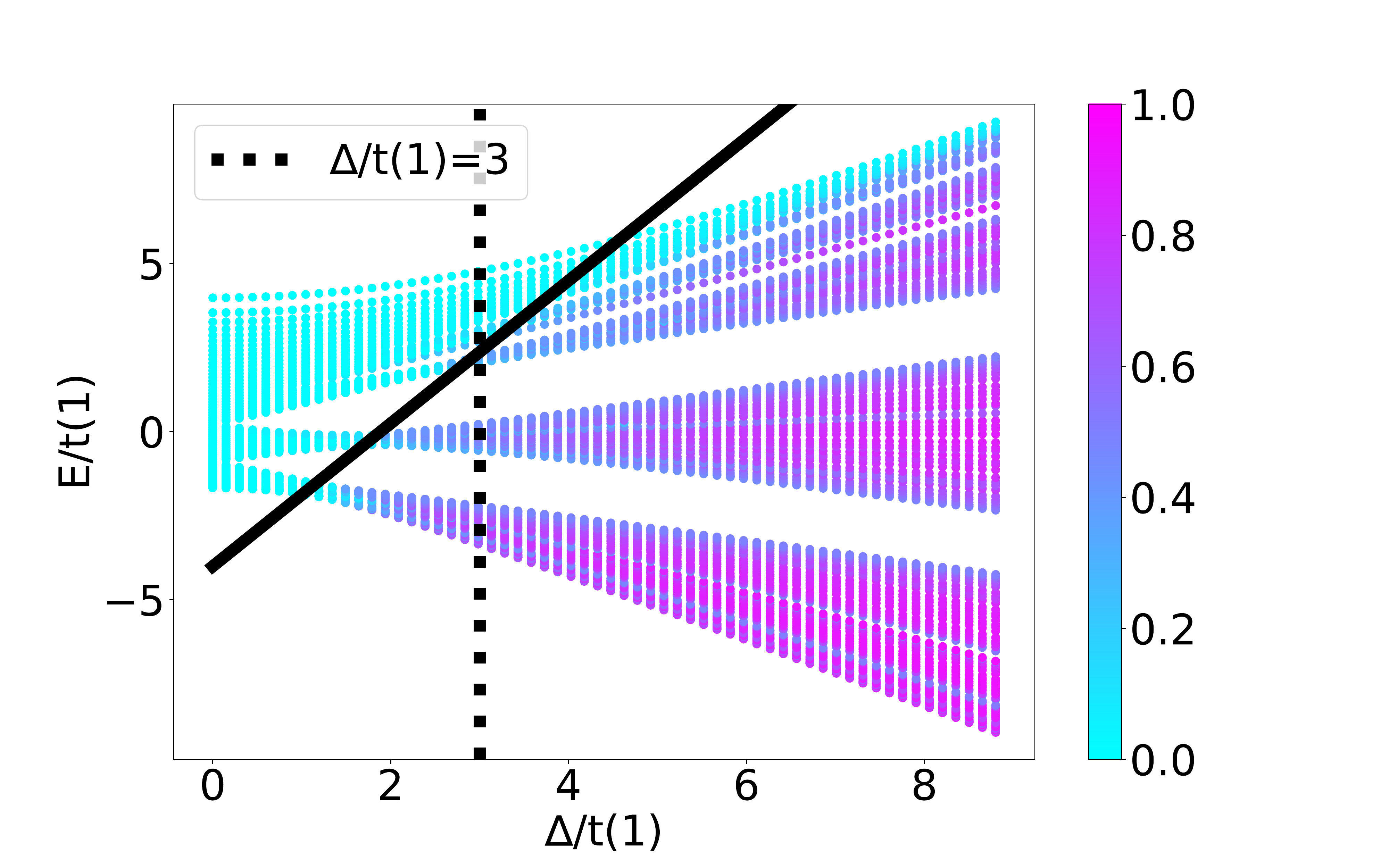}
    \caption{(Upper panel) Contour plot of Energy dependent IPR vs $\Delta/t(1)$ for the Hamiltonian $H$. Solid lines show analytical separation between delocalized and localized states.
    (Lower panel) Contour plot of Energy dependent IPR vs $\Delta/t(1)$ for the Hamiltonian $H_{\text{eff}}$ for $q=1.528$. Solid lines show analytical separation between delocalized and localized states. 
    Results are for $\alpha=0.5$ and $L=160$.}
    \label{fig5a}
\end{figure}

\begin{figure}
    \centering
    \includegraphics[width=0.53\textwidth, height=0.3\textwidth]{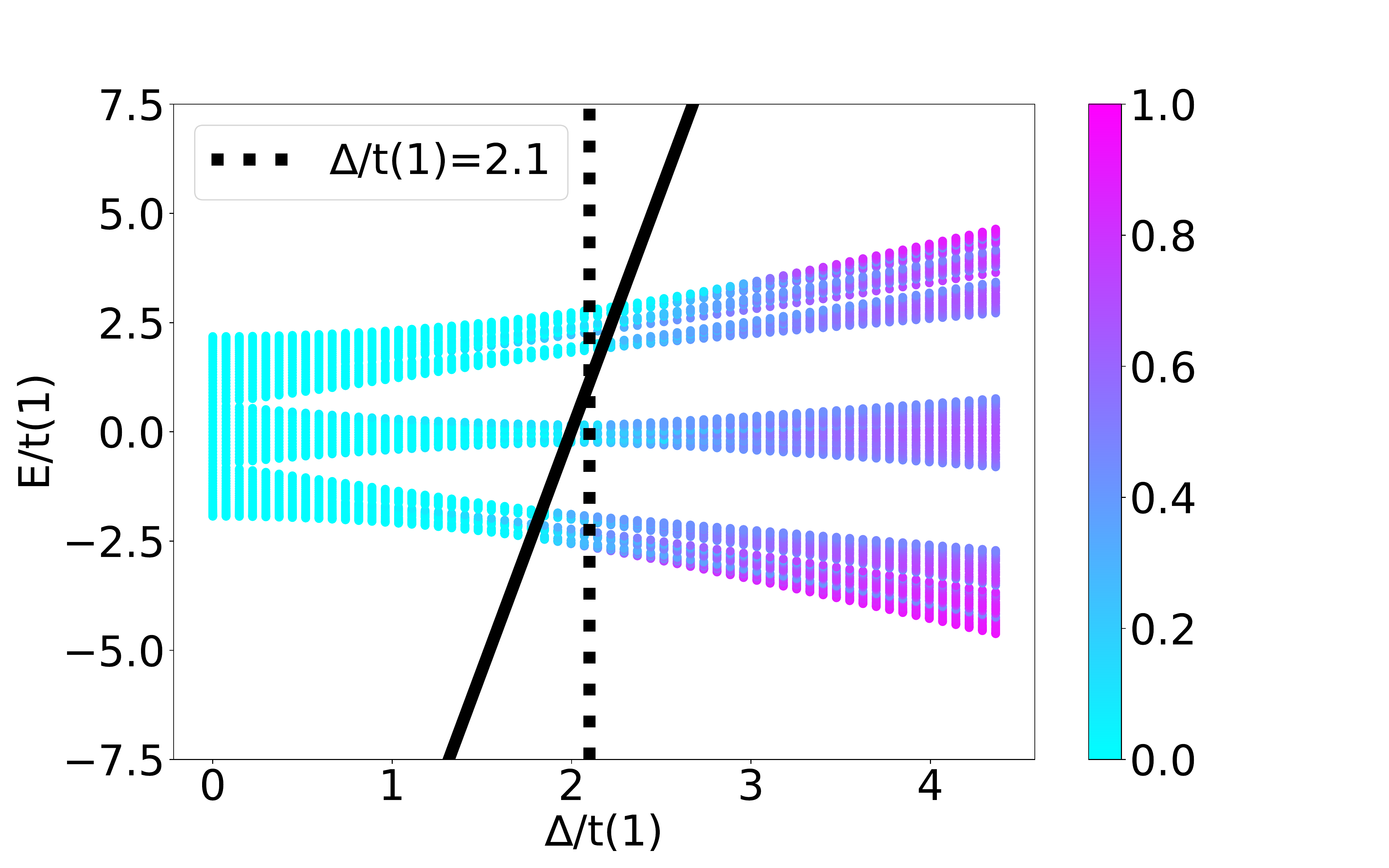}
    \includegraphics[width=0.53\textwidth, height=0.3\textwidth]{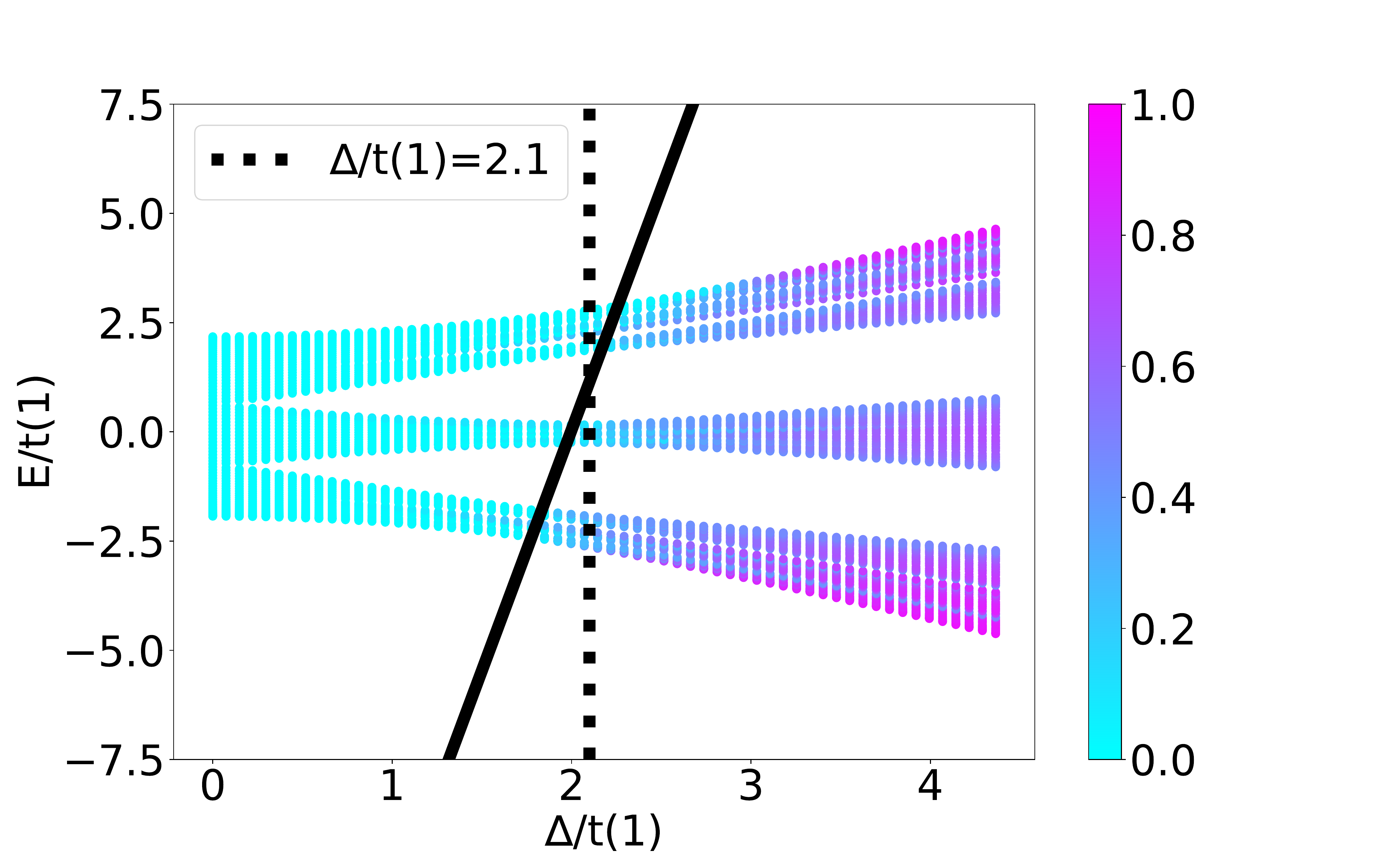}
    \caption{(Upper panel) Contour plot of Energy dependent IPR vs $\Delta/t(1)$ for the Hamiltonian $H$. Solid lines show analytical separation between delocalized and localized states.
    (Lower panel) Contour plot of Energy dependent IPR vs $\Delta/t(1)$ for the Hamiltonian $H_{\text{eff}}$ for $q=2.4689$. Solid lines show analytical separation between delocalized and localized states. 
    Results are for $\alpha=1.5$ and $L=160$.}
    \label{fig5b}
\end{figure}

\section{Effective Hamiltonian}
\textcolor{black}{From our numerical study it is obvious that the phase diagram for L\'evy quasicrystal is very similar to the long-range hopping model studied in Ref.~\cite{even-odd-levelspacingPhysRevLett.123.025301}. We study the variation of $t(n)/t(1)$ vs $n$ in the upper panel of Fig.~\ref{fig3}. We find that 
the fall-off of hopping amplitude for the range $n\ge2$, 
$t(n)\sim t(2)/|n/2|^{q}$, where  $q$ depends on $\alpha$}. Hence, we write down an effective Hamiltonian, 
\begin{eqnarray}
H_{\text{eff}}&=&\sum_{j}t(1)({c}^{\dag}_j{c}_{j+1}+\text{H.c.})+\sum_{j,n>1}\frac{t(2)}{|n/2|^{q}}({c}^{\dag}_j{c}_{j+n} \nonumber\\
&&+\text{H.c.})
+\Delta\sum_{j} \cos(2\pi\beta j+\phi){n}_j,
\label{eff_model}
\end{eqnarray}
and compare it with L\'evy quasicrystal. 
In the large $n$ limit,  it can be shown analytically that $q\simeq \alpha+1$ (see Appendix: \ref{appendixIV}), and the $q$
obtained from the numerical calculations (see inset of the lower panel of Fig.~\ref{fig3}) also show a similar result. 
The structure of this effective Hamiltonian is quite similar to the previously  studied AA models with power-law hopping in Ref:~\cite{even-odd-levelspacingPhysRevLett.123.025301,sharma.2021} except that in our case we have an extra nearest-neighbor 
hooping term.  \textcolor{black}{We like to emphasize that while Fig.~\ref{fig3} (upper panel) might suggest that $H_{eff}$ is an extremely accurate description of the L\'evy quasicrystal, a careful study suggests that there are significant differences. In the inset, we show the variation of the relative difference in hopping amplitudes between these two models, which is defined as
\begin{eqnarray}
\delta(n)=\frac{|t^{H}(n)-t^{H_{eff}}(n)|}{t^{H}(n)},  
\end{eqnarray}
where $t^H$ and $t^{H_{eff}}$ stands for hopping amplitudes of L\'evy
 quasicrystal and effective power-law hopping Hamiltonian respectively. 
The relative difference is more than $20\%$ for some values of $\alpha$ (see inset of Fig.~\ref{fig3} upper panel). 
}
Moreover, we also calculate the normalized trace distance~\cite{modak.2017} between the Hamiltonian $H$ \eqref{nonint_model} and $H_{\text{eff}}$.  
The normalized trace distance is defined as follows, 
$d(H,H_{\text{eff}})=\frac{1}{L}\text{Tr}[\sqrt{(H-H_{\text{eff}})^2}]$.
Lower panel of Figure.~\ref{fig3} shows the variation of $d(H,H_{\text{eff}})$ with $\alpha$, this trace difference between two Hamiltonians are small $<10^{-2}$, but not negligible. Hence, one can conclude that $H_{\text{eff}}$ is a reasonably good approximation of our original Hamiltonian $H$, \textcolor{black}{but they still have significant differences.}


\textcolor{black}{Hence, it is not obvious that both of these Hamiltonians will show a similar phase diagram.}
Next, we compare the energy-dependent IPR for both the Hamiltonians($H$ and $H_{\text{eff}}$) and remarkably find that they are almost qualitatively very similar. 
We show the results for $\alpha=0.5$ and $1.5$ and compare them with $H_{\text{eff}}$ results. The results of the energy-dependent IPR are shown in Fig.~\ref{fig5a} and Fig.~\ref{fig5b} respectively. The solid line represents the ME line obtained using Eqn.~\ref{duality condition1}.  It seems that 
the analytic prediction of the ME line does qualitatively a very good job for $\alpha=1.5$ compared to $\alpha=0.5$, presumably because $t(2)/t(1)$ is much smaller for $\alpha=1.5$.
It's in great agreement with our numerical prediction. This gives us the validation that indeed $H_{\text{eff}}$ is a good approximation to our original Hamiltonian $H$ even though they are not identical(see Appendix.~\ref{appendix0}).\\
Also, this effective Hamiltonian can be experimentally probed for polar molecules pinned in deep bichromatic optical lattices. Powers $1 \le q \le 3$ (which is also what we have obtained for our model) may be directly realized in ions~\cite{richerme2014non,jurcevic2014quasiparticle}.

\section{Conclusions}
In this work, we define L{\'e}vy quasicrystal by discretizing the space-fractional Schr$\ddot{\text{o}}$dinger equation using
the Gr$\ddot{\text{u}}$nwald derivatives and adding on-site quasiperiodic potential and investigate the phase diagram. We find that while there exists a parameter region where all states are either completely delocalized or
completely localized \textcolor{black}{as one will observe for normal quasicrystal}, there also exists a region where delocalized and localized states coexist and are separated by a mobility edge. Moreover, based on the fraction of DL states,
we identify different $p_s$ ME phases. 
\textcolor{black}{Similar phases have been observed in the AA model with power-law hopping~\cite{even-odd-levelspacingPhysRevLett.123.025301,sharma.2021}. Hence, we compare our results with the effective power-law hopping model as well. To the best of our knowledge such $p_s$ ME phases have not been observed in any other models so far except for the power-law hopping models. Our results prove that similar $p_s$ phases can be found even for models where the fall-off of hopping amplitudes are not strictly power-law type. It opens up a broader question, in order to have $p_s$ ME phases, what necessary condition one needs to impose on hopping amplitude? Our study also shows a connection between $p_s$ ME phases with SFQM, which might be an important step forward in understanding the origins of such $p_s$ ME phases. 
}
\textcolor{black}{
It is well established that to understand the traveling behavior of animals~\cite{brockmann2006scaling,benhamou2007many}, the complex dynamics of real-life financial markets~\cite{yarahmadi20222d}, L{\'e}vy description has been extremely useful. If one wants to model a situation to understand how these things can get affected due to correlated random events, our L\'evy quasicrystal will be an automatic choice. One of the biggest advantages of  L\'evy quasicrystal is that apart from the localized and delocalized phases; it also supports a mixed phase ( mobility edges that separate localized and delocalized states). 
This can be thought of as those random correlated events that affect the motions of some animals while the motions of other animals are unaffected. 
On top of that, the existence of different $p_s$ mobility edge phases (which is unique for our model) gives us extreme control over the ratio of 
the number of animals whose motion gets restricted (localized states) and the number of animals that can still move freely (delocalized states). Note that one can get all this control in our model with only two free parameters, i.e., $\alpha$ (Levy index) and $\Delta$ (strength of correlated random potential). }

\textcolor{black}{Moreover, in the optical system, there has been a proposal to experimentally realize a very similar model~\cite{huang2019localization}, we strongly believe our study should motivate experimentalists to revisit the same model and identify such $p_s$ mobility edge phases.}
 We also think that given there are similarities in the phase diagram between L{\'e}vy quasicrystal and power-law hopping models, it makes L{\'e}vy quasicrystal a
probable candidate for the experimentally accessible realization of space-fractional quantum mechanics (SFQM) in a condensed-matter set-up. Further, it will be interesting to investigate the role of interaction ~\cite{modak.2015,deng2017many,arti.2017} and the effect of the Hermiticity breaking and the $PT$ symmetry  ~\cite{nonh_qp,modak.2021,shukla2022heisenberg} in the context of SFQM.

\section{Acknowledgements}
RM acknowledges the DST-Inspire research grant by the Department of Science and Technology, Government of India, SERB start-up grant (SRG/2021/002152). The authors thank Vikash Pandey
and Bhabani Prasad Mandal for introducing the topic of 
Space-fractional quantum mechanics. Authors also thank 
Sambuddha Sanyal for fruitful discussions.

\appendix

\begin{figure}
    \centering
    \includegraphics[width=0.48\textwidth, height=0.3\textwidth]{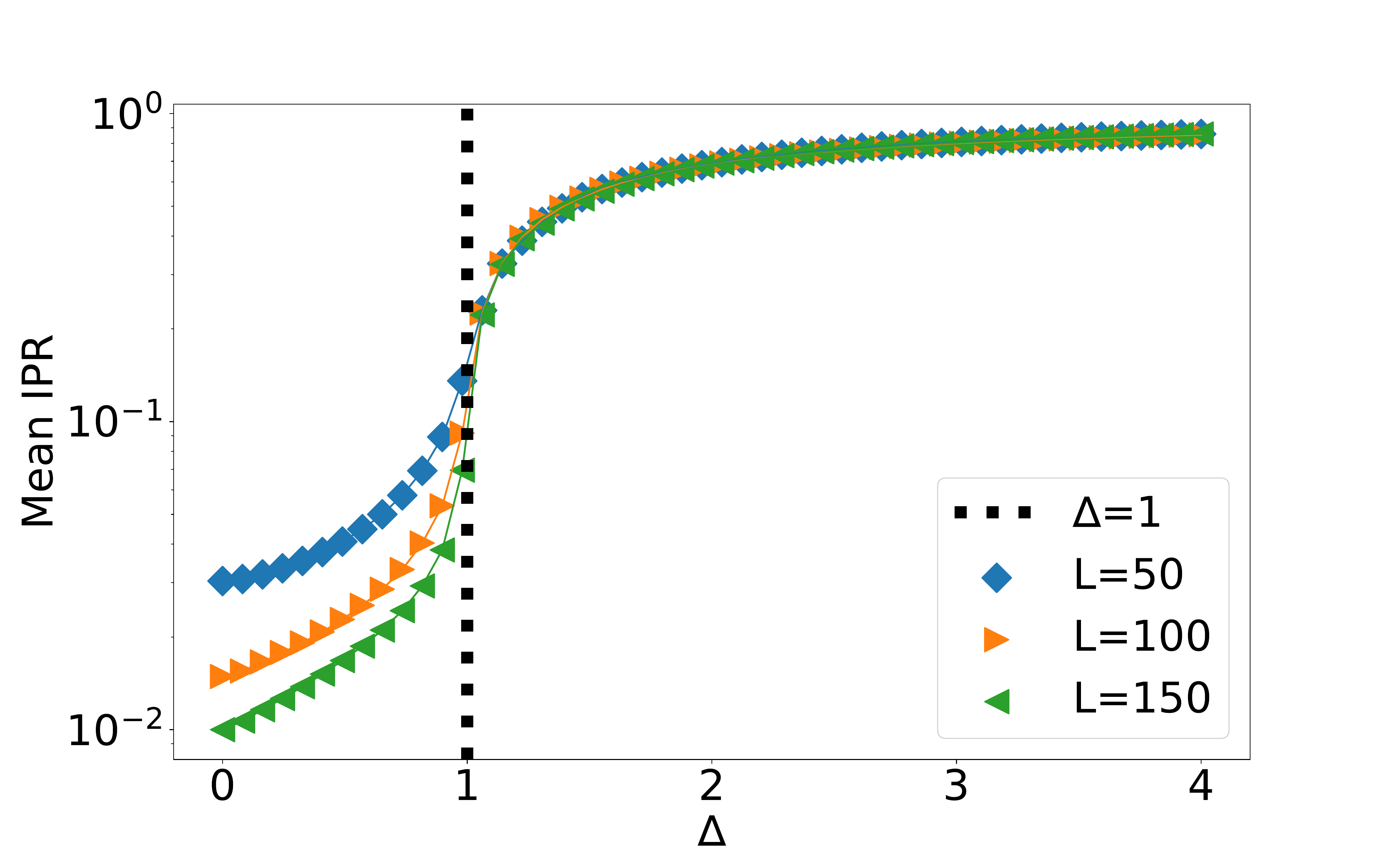}
    \caption{Variation of Mean IPR vs $\Delta$ for the $\alpha=1$ model Eqn.~\ref{nonint_model_a1}. The vertical dashed line shows the transition i.e. $\Delta=1$. }
    \label{fig7}
\end{figure}

\begin{figure}
    \centering
    \includegraphics[width=0.48\textwidth, height=0.3\textwidth]{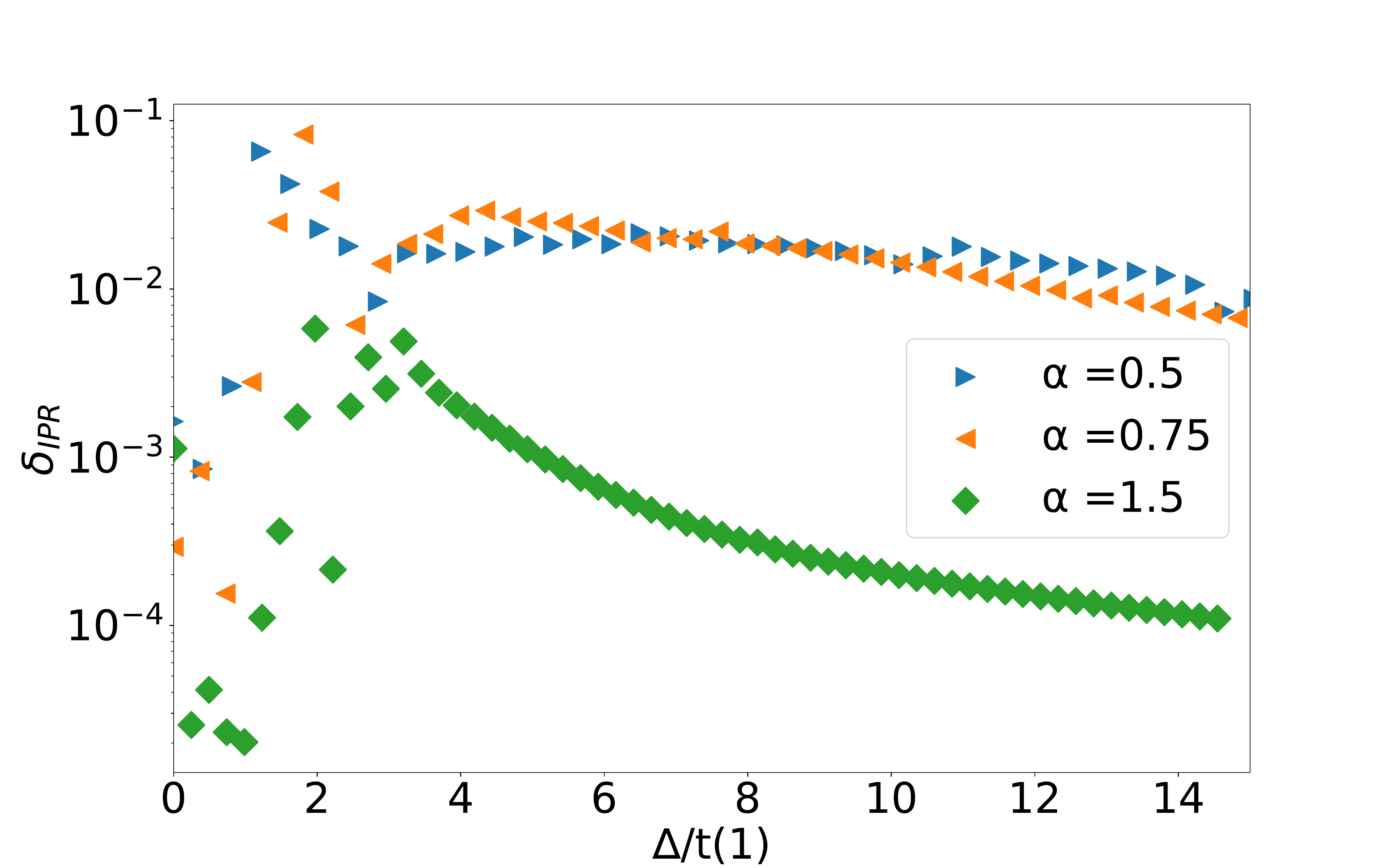}
    \caption{Variation of relative difference of mean IPR between $H$ and $H_{eff}$ vs $\Delta/t(1)$ for different values of $\alpha$.}
    \label{fig10}
\end{figure}

\section {$\alpha=1$ Model} \label{appendixII}
In the main text, we introduce L{\'e}vy quasicrystal by discretizing the space-fractional Schr$\ddot{\text{o}}$dinger equation using the Gr$\ddot{\text{u}}$nwald-Letnikov derivatives and adding on-site quasiperiodic potential. However, this 
discretization technique works for $\alpha\in (0,2]$ except 
$\alpha=1$. Hence, we focus on the $\alpha=1$ model separately. 

For $\alpha=1$, $\langle x_l|D_{\alpha}P^{\alpha}|\psi \rangle$ becomes, 
\begin{eqnarray}
\langle x_l|D_{\alpha}P^{\alpha}|\psi\rangle=\frac{i}{2a}[\psi(x_l+a)-\psi(x_l-a)], 
\end{eqnarray}
where $a$ is the lattice constant. 
The Hamiltonian $H$ is given by, 
\begin{eqnarray}
H&=&\sum_{j}\frac{1}{2}(i{c}^{\dag}_j{c}_{j+1}+\text{H.c.})+\Delta\sum_{j} \cos(2\pi\beta j+\phi){n}_j. \nonumber \\
\label{nonint_model_a1}
\end{eqnarray}
Note that for $\Delta=0$, the energy dispersion relation of the  Hamiltonian is $E(k)=\sin k$, in the continuum limit i.e. $k\to0$, $E(k)\sim k$ (linear dispersion model). 
This model shows localization-delocalization transition 
at $\Delta=1$. We have shown this by calculating Mean IPR 
as a function of $\Delta$ in Fig.~\ref{fig7}. 

\begin{figure}
    \centering
    \includegraphics[width=0.48\textwidth, height=0.3\textwidth]{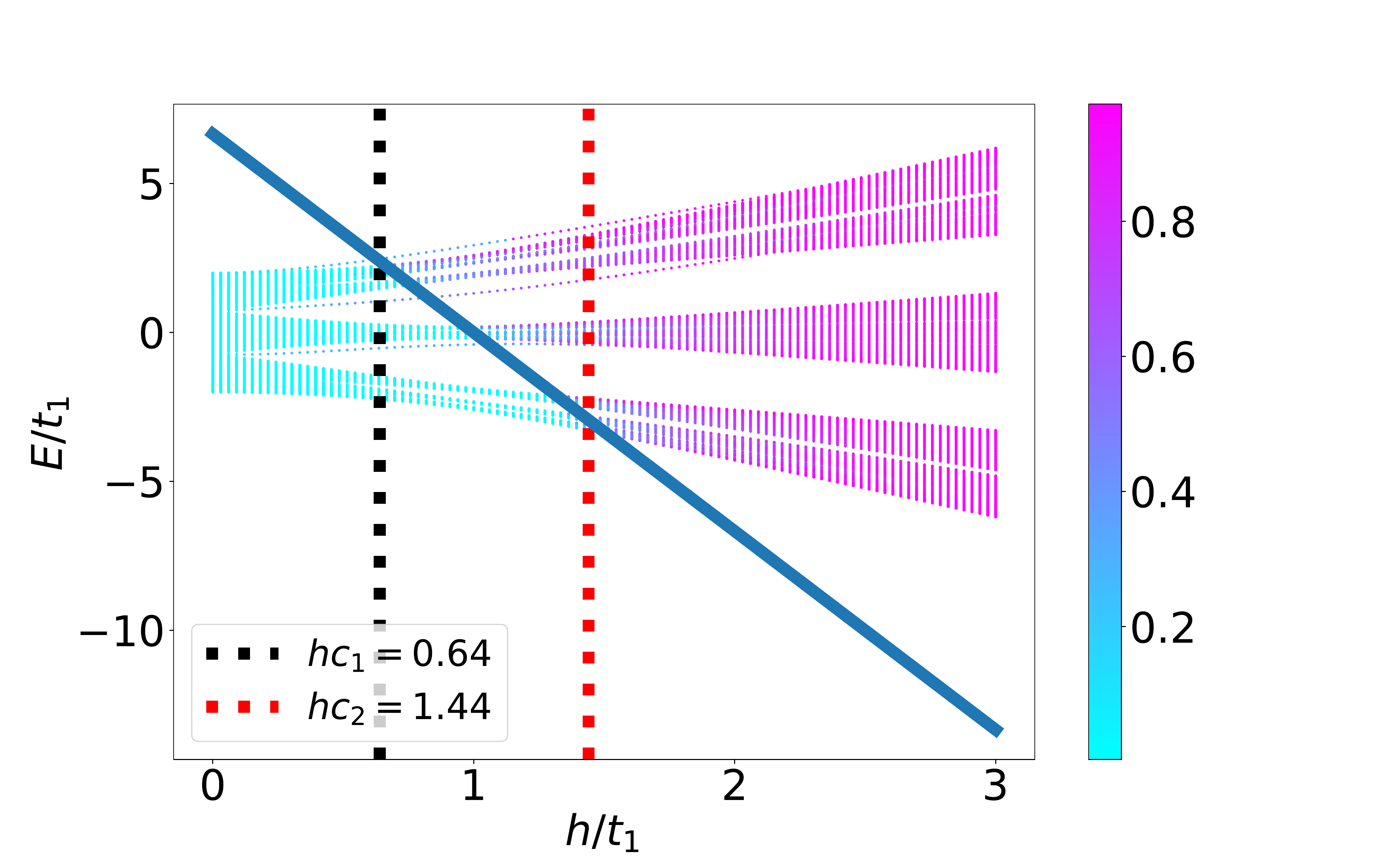}
    \caption{Contour plot of energy dependent IPR vs $\frac{h}{t_1}$ for the GAA hamiltonian\eqref{GAA Hamiltonian}.The solid blue line shows the exact separation of delocalized and localized states, given by Eqn.\eqref{mobility edge line of GAA}. Dashed black and red lines show the transition points calculated from the tolerance method.Here we take $\gamma=0.3$ and $t_1=1$}
    \label{fig8}
\end{figure}

\section{Comparison between Mean IPR of $H$ and $H_{eff}$}
\label{appendix0}
In the main text, we have shown results for energy-dependent IPR for L\'evy quasicrystal and $H_{eff}$. While these two models are not identical but the IPR results look qualitatively very similar (see Fig.~\ref{fig5a} and Fig.~\ref{fig5b}). We define the relative difference of mean IPR ($M$) between $H$ and $H_{eff}$ as,
\begin{eqnarray}
        \delta_{IPR} =\frac{|M_{H}-M_{H_{eff}}|}{M_{H}} .
\end{eqnarray}
Figure.~\ref{fig10} shows the variation $\delta_{IPR}$ vs $\Delta/t(1)$ for different values of $\alpha$, and we find that the relative difference is less than $10\%$. Note that on the other hand,  the relative difference in hopping amplitude between these two models is almost $20\%$ as discussed in the main text.

\section {Method for detecting different phases for finite system} \label{appendixIII}
While IPR is an extremely useful diagnostic to detect localized and delocalized states, in order to really identify DL and AL states, one needs to do a finite-size scaling analysis. Given our Hamiltonian $H$ involves terms having Gamma function in the hopping parameter, numerically it is very difficult to study large systems, especially beyond $L=200$ (long-range hopping terms can't be calculated accurately). Hence, in this work, we are restricted to small system sizes, and doing a proper finite size analysis is tricky. In order to understand the delocalization-localization transition, we use tolerance criteria based on the finite-size AA model. 
We calculate the difference between mean PR for the AA model 
(for the Hamiltonian $H$ i.e. $\alpha=2$ and $\Delta=2$)
for two system sizes $L_1$ and $L_2$ at the transition point  i.e. $\epsilon=PR(L_1)-PR(L_2)$. We use that value to detect different phases. 

First, we use this criteria for the Generalized Aubry-Andr\'{e} model, which is given by,\cite{deng2017many}\\
\begin{eqnarray}
\tilde{H}&=&\sum_{j}t_1({c}^{\dag}_j{c}_{j+1}+\text{H.c.})+2h\sum_{j} \frac{\cos(2\pi\beta j+\phi)}{1-\gamma\cos(2\pi\beta j+\phi)}{n}_j. \nonumber \\
\label{GAA Hamiltonian}
\end{eqnarray}
where $\gamma \in(-1,1)$ .
This model is having an exact analytical solution for the mobility edge line as,
\begin{eqnarray}
\gamma E=2\text{sgn}(h)(|t_1|-|h|). 
\label{mobility edge line of GAA}
\end{eqnarray}

\begin{figure}
    \centering
    \includegraphics[width=0.53\textwidth, height=0.3\textwidth]{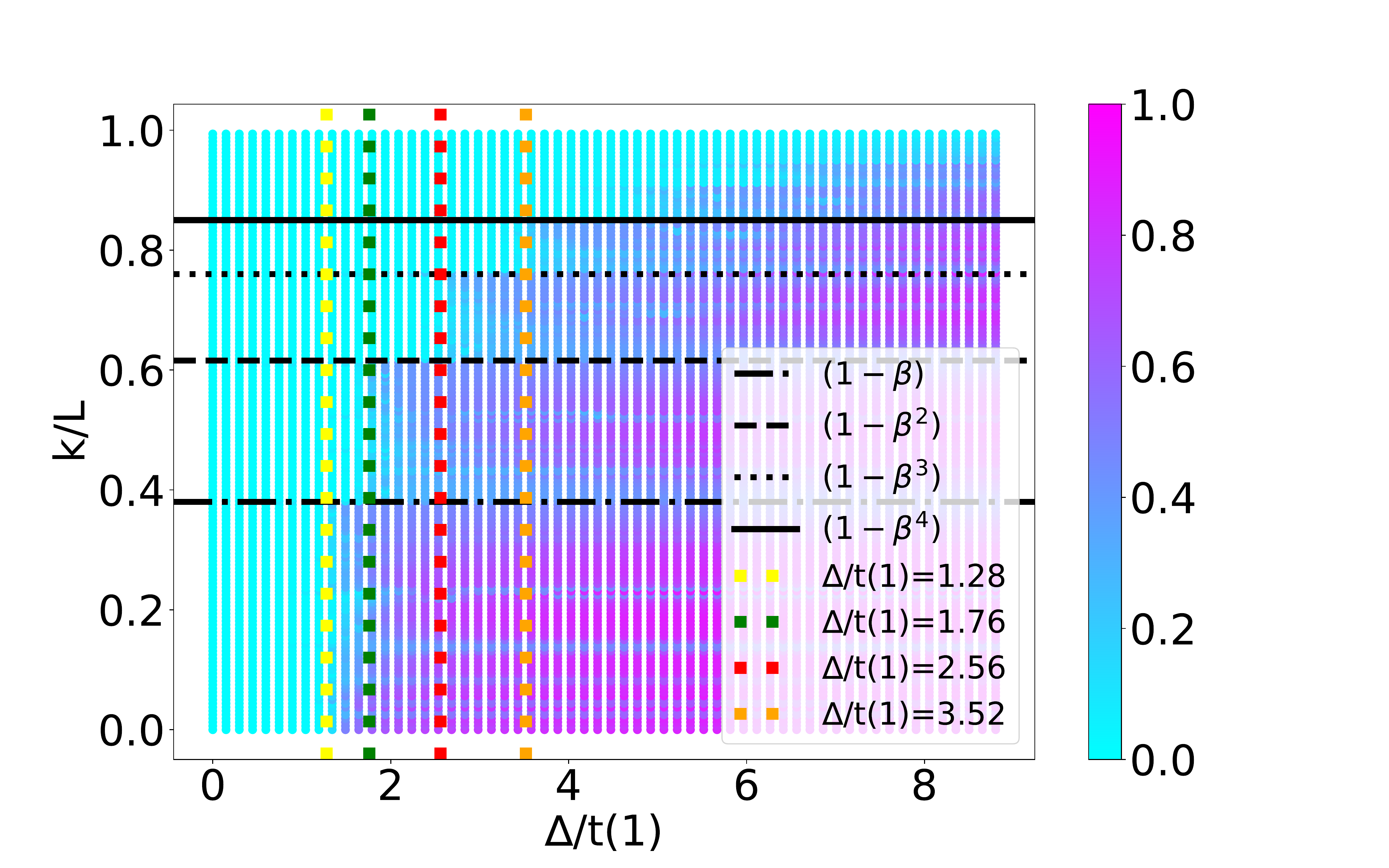}\\
    \caption{Shows the 
    variation of IPR with respect to energy level index $k/L$. 
    for $\alpha=0.5$. Vertical lines are phase boundaries between different $p_s$ phases obtained from the tolerance criteria.}
    \label{fig9}
\end{figure}

We use our tolerance criteria on the ground state and the highest excited state for the same $L_1$ and $L_2$ for which the tolerance criteria for the AA model have been obtained, to identify $hc_1$ (below which all states are DL) and $hc_2$ (above which all states are AL).
They are in great agreement with the exact analytical ME prediction as shown in Fig.~\eqref{fig8}. 

That motivates us to go forward and use the same criteria 
to detect different $p_s$ ME phases in the $H$. We calculate 
 PR for the Hamiltonian $H$ for $m$th  eigenstate ($m$ is the nearest integer of $(1-\beta^s)L-1$) for different values of $\Delta$, whenever they match with our tolerance criteria, we identify that value of $\Delta$ as a transition point between $p_{s-1}$ ME phase to $p_s$ ME phase ($p_0$ phase can be thought of DL phase). Figure.~\ref{fig9}
clearly demonstrate that our criteria do a reasonably 
good job of identifying those phases. We use this technique to obtain the phase diagram which is presented in the main text. We have repeated our calculation for large systems for $H_{\text{eff}}$ and obtained a very similar phase diagram as well.
\section{Analytical prediction: the equivalence between $H$ and $H_{eff}$ in large $n$ limit}\label{appendixIV}
Using the reflection property of Gamma function one can write,\\
$\Gamma(z)\Gamma(1-z)=\frac{\pi}{\sin(z\pi)}$\\
Moreover,\\
$\Gamma(|n|-\alpha)\Gamma(1-|n|+\alpha)=\frac{\pi}{\sin[(|n|-\alpha)\pi]}=\frac{\pi (-1)^{(|n|-1)}}{\sin(\pi\alpha)}$\\
Now, $\sin(\alpha\pi)=\frac{\pi}{\Gamma(-1-\alpha)\Gamma(2+\alpha)}$, substituting this value to the above equation we can write
\begin{eqnarray}
\label{D1}
    \Gamma(1-|n|+\alpha)=\frac{(-1)^{|n|-1}\Gamma(-1-\alpha)\Gamma(\alpha+2)}{\Gamma(|n|-\alpha)}
\end{eqnarray}

Now for $0<\alpha<1$ limit,\\
$t(n)=-\frac{1}{2}[(-1)^{{|n|}}\binom{\alpha}{|n|}]$\\
$~~~~~~=-\frac{1}{2}\frac{(-1)^{|n|} \Gamma(\alpha+1)}{\Gamma(|n|+1)\Gamma(\alpha-|n|+1)}$\\

Now taking $n\to \infty$ limit and using Stirling’s formula($\Gamma(|n|-\alpha)\approx \sqrt{2\pi}(|n|-\alpha)^{{(|n|-\alpha)-\frac{1}{2}}}e^{-(|n|-\alpha)}$ and $\Gamma(|n|+1)\approx \sqrt{2\pi}(|n|+1)^{{(|n|+1)-\frac{1}{2}}}e^{-(|n|+1)}$ ), and using Eqn:\eqref{D1}, $t(n)$ becomes,
\begin{eqnarray}
    t(n)\simeq\frac{1}{2} \frac{e^{\alpha-1}}{(\alpha+1)\Gamma(-1-\alpha)}\frac{1}{|n|^{\alpha+1}},~~ \text{for}~~ 0<\alpha<1~~~\nonumber \\
\end{eqnarray}
Similarly using the reflection property of the Gamma function one can write,\\
\begin{eqnarray}
    \Gamma(\alpha-|n|)&=&\frac{-(-1)^{|n|-1}\pi}{\Gamma(|n|-\alpha+1)\sin(\alpha \pi)}\nonumber\\
\text{or},\Gamma(\alpha-|n|)&=&\frac{-(-1)^{|n|-1}\Gamma(-1-\alpha)\Gamma(\alpha+2)}{\Gamma(|n|-\alpha+1)}
\label{D3}~~~
\end{eqnarray}
For the limit $1<\alpha<2$,

   $ t(n)=\frac{1}{2}[(-1)^{|n|+1}\binom{\alpha}{|n|+1}] $ \\
   $~~~~~~~~~=\frac{1}{2}\frac{(-1)^{|n|+1} \Gamma(\alpha+1)}{\Gamma(|n|+2)\Gamma(\alpha-|n|)}$\\
   
Taking $n\to \infty$ limit and using Stirling’s formula($\Gamma(|n|-\alpha+1)\approx \sqrt{2\pi}(|n|-\alpha+1)^{(|n|-\alpha+1)-\frac{1}{2}}e^{-(|n|-\alpha+1)}$ and $\Gamma(|n|+2)\approx \sqrt{2\pi}(|n|+2)^{(|n|+2)-\frac{1}{2}}e^{-(|n|+2)}$ ), and using Eqn:\eqref{D3}, $t(n)$ becomes,
\begin{eqnarray}
    t(n)\simeq-\frac{1}{2} \frac{e^{\alpha+1}}{(\alpha+1)\Gamma(-1-\alpha)}\frac{1}{|n|^{\alpha+1}},~ \text{for}~ 1<\alpha<2~~~~\nonumber\\
\end{eqnarray}
Hence, indeed in the large $n$ limit, the hopping parameter $t(n)$ shows power law behavior.
\bibliography{cite}
\end{document}